\documentclass[aps,prb,12pt,onecolumn,showpacs,superscriptaddress]{revtex4-2}

\usepackage{soul}
\usepackage[colorlinks=true, linkcolor=blue, anchorcolor=blue, citecolor=blue, urlcolor=blue]{hyperref}
\usepackage{amssymb,graphicx,color,amsmath,mathtools,physics}
\usepackage{xfrac,bm,stmaryrd,trimclip,xcolor}
\usepackage[mathletters]{ucs}
\usepackage{gensymb}

\newcommand{\CVS}{CsV$_3$Sb$_5$}
\newcommand{\AVS}{\textit{A}V$_3$Sb$_5$}

\begin{document}
    
\title{Probing orbital magnetism of a kagome metal \CVS\ by a tuning fork resonator}

\author{Hengrui Gui}
\thanks{These authors contributed equally to this work.}
\affiliation{Center for Correlated Matter and School of Physics, Zhejiang University, Hangzhou 310058, China}

\author{Lin Yang}
\thanks{These authors contributed equally to this work.}
\affiliation{College of Materials and Environmental Engineering, Hangzhou Dianzi University, Hangzhou 310018, China}

\author{Xiaoyu Wang}
\email[Corresponding author: ]{wxybkaphsc@gmail.com}
\affiliation{National High Magnetic Field Laboratory, Florida State University, Tallahassee, FL 32310, USA}

\author{Dong Chen}
\affiliation{Max Planck Institute for Chemical Physics of Solids, 01187 Dresden, Germany}
\affiliation{College of Physics, Qingdao University, Qingdao 266071, China}

\author{Zekai Shi}
\affiliation{Center for Correlated Matter and School of Physics, Zhejiang University, Hangzhou 310058, China}

\author{Jiawen Zhang}
\affiliation{Center for Correlated Matter and School of Physics, Zhejiang University, Hangzhou 310058, China}


\author{Jia Wei}
\affiliation{Center for Correlated Matter and School of Physics, Zhejiang University, Hangzhou 310058, China}

\author{Keyi Zhou}
\affiliation{Center for Correlated Matter and School of Physics, Zhejiang University, Hangzhou 310058, China}

\author{Walter Schnelle}
\affiliation{Max Planck Institute for Chemical Physics of Solids, 01187 Dresden, Germany}

\author{Yongjun Zhang}
\affiliation{Hubei Key Laboratory of Photoelectric Materials and Devices, School of Materials Science and Engineering, Hubei Normal University, Huangshi, China}

\author{Yu Liu}
\affiliation{Center for Correlated Matter and School of Physics, Zhejiang University, Hangzhou 310058, China}

\author{Alimamy F. Bangura}
\affiliation{National High Magnetic Field Laboratory, Florida State University, Tallahassee, FL 32310, USA}

\author{Ziqiang Wang}
\affiliation{Department of Physics, Boston College, Chestnut Hill, MA 02467, USA}

\author{Claudia Felser}
\affiliation{Max Planck Institute for Chemical Physics of Solids, 01187 Dresden, Germany}

\author{Huiqiu Yuan}
\email[Corresponding author: ]{hqy@zju.edu.cn}
\affiliation{Center for Correlated Matter and School of Physics, Zhejiang University, Hangzhou 310058, China}
\affiliation {Institute for Advanced Study in Physics, Zhejiang University, Hangzhou 310058, China}
\affiliation {Institute of Fundamental and Transdisciplinary Research, Zhejiang University, Hangzhou 310058, China}
\affiliation  {State Key Laboratory of Silicon and Advanced Semiconductor Materials, Zhejiang University, Hangzhou 310058, China}

\author{Lin Jiao}
\email[Corresponding author: ]{lin.jiao@zju.edu.cn}
\affiliation{Center for Correlated Matter and School of Physics, Zhejiang University, Hangzhou 310058, China}
\maketitle
\cleardoublepage

\textbf{
The recently discovered kagome metal \CVS\ exhibits a complex phase diagram  that encompasses frustrated magnetism, topological charge density wave (CDW), and superconductivity. One CDW state that breaks time-reversal symmetry was proposed in this compound, while the exact nature of the putative magnetic state remains elusive.
To examine the thermodynamic state of \CVS\ and assess the character of the associated magnetism, we conducted tuning fork resonator measurements of magnetotropic susceptibility over a broad range of angles, magnetic fields, and temperature.
We found a cascade of phase transition in the CDW phase. Of particular interest is a highly anisotropic magnetic structure that arises below about 30~K, with a magnetic moment along the $c$-axis that has an extremely small magnitude.
This magnetic state demonstrates extremely slow dynamics and small saturate field, all suggest that electronic phase below 30~K breaks time reversal symmetry and has an unconventional origin.
}

\section{Introduction}

Materials with a kagome lattice have gathered significant research interest recently~\cite{CheckelskyJ2018, FelserC2018a, Hasan2018}. The interplay between band topology and electronic correlations in such systems could give rise to novel electronic phases and exotic quasiparticles~\cite{Hasan2022Review}. 
\AVS ($A$ = K, Cs, or Rb) hosts kagome lattice of vanadium ions stacked between antimony and alkali ion layers (Fig.~\ref{fig1}(a)).
Around 80-100~K, \AVS\ exhibits a well-characterized CDW phase ~\cite{TobererES2019, WilsonS2020, LeiHC2021CPL,Tan2021}, followed by a series of symmetry breaking transitions upon lowering the temperature. These include rotational~\cite{WenHH2021NC, ChenXH2022N, Ilija2022NP, ilija2023NP} or mirror~\cite{Hasan2021, Vidya_optical_RbVSb} symmetry breaking, as well as superconducting phase transition~\cite{WilsonS2020, LeiHC2021CPL}.
Furthermore, a time-reversal breaking phase has been proposed based on scanning tunneling spectroscopy (STM)~\cite{Hasan2021}, muon spin resonance ($\mu$SR)~\cite{ZhaoZX2021, Guguchia2022}, non-linear resistivity~\cite{MollP2022}, and Kerr rotation~\cite{WuL2022} measurements, while many other groups did not find any signature of time-reversal breaking using the same probes~\cite{ChenXH2022N, Ilija2022NP, ilija2023NP, WenHH2022PRB, Kapitulnik2022}.
However, no static magnetic moment or associated long-range magnetic order could be resolved by several advanced techniques~\cite{TobererES2019, ZhaoZX2021, GrafMJ2021, Guguchia2022, Mike2022, WilsonS2020}. Therefore, intensive debates arise regarding the nature of the low-temperature phase in \AVS, especially in the case of \CVS~\cite{GuoNP2024}, and the boundaries between different phases remain elusive. 

It has been suggested that the time-reversal symmetry breaking phase in \CVS\ is unconventional and has an orbital origin~\cite{HuJP2021, Balents2021,NeupertT2021, Nandkishore2021,HuJP2021PRB,Zhou2022,ChristensenPRB2022,Tazai2022,Dong2023}. This phase is akin to the Haldane flux states on honeycomb lattice~\cite{Haldane1988} or the loop currents in charge-transfer systems~\cite{Giamarchi1995, Varma1997, Arkady2022PRB}. In the \AVS\ family, these orbital phases are currently discussed as ``loop current CDW'', or ``CDW with loop current order''~\cite{Hasan2021, MollP2022, ZhaoZX2021}. This state provides a novel origin of magnetism. What is missing is a thermodynamic signature of such orbital magnetism in \CVS.
Theoretically, such orbital magnetic states can produce very small and, naturally, strongly anisotropic magnetic moments when supported by a large number of transition metal ions~\cite{Varma1997, Arkady2022PRB}. This is because orbital currents flow along adjacent sub-lattice, generating local moments strictly along the out-of-plane direction. 
In contrast, the localized spin-magnetism~\cite{Anderson1961} in a 3d transition metal system (which has small spin-orbit coupling~\cite{Rado, Greenwood1984}) that tends to be weakly anisotropic. Therefore, examining the magnetism and its anisotropy is crucial for uncovering the nature of the low-temperature phase in \CVS. 

Recently, a quartz tuning fork based technique was developed and can detect small anisotropy of susceptibility~\cite{Arkady2021}.
The change in the tuning fork resonant frequency $\Delta f$ is directly proportional to the so-called magnetotropic susceptibility ($k(\theta)$)~\cite{Arkady2023}, which is defined as
$k$ = $\partial^2F(\mathbf{B})/\partial \theta^2_{\mathbf{n}}$ $\propto$ $\Delta f(\theta)$, 
where $\mathbf{n}$ is the axis of rotation, $F$ is the free energy, $\theta$ is the angle between $\textbf{B}$ and the sample (Note: we use $\textbf{B}$ to replace $\textbf{H}$ for convenience, as the overall susceptibility of \CVS\ is small. See Supplementary Note 1 and Supplementary Figs. 1 and 2 for working principles of the tuning fork resonator). In this work, we show extensive tuning fork resonator measurements of the magnetotropic susceptibility in a broad temperature- and magnetic field range. Our results provide direct evidence for a small but highly anisotropic magnetic moment in \CVS.

\section{Results}
\begin{figure*}[t!!]
\includegraphics[width=.85\textwidth]{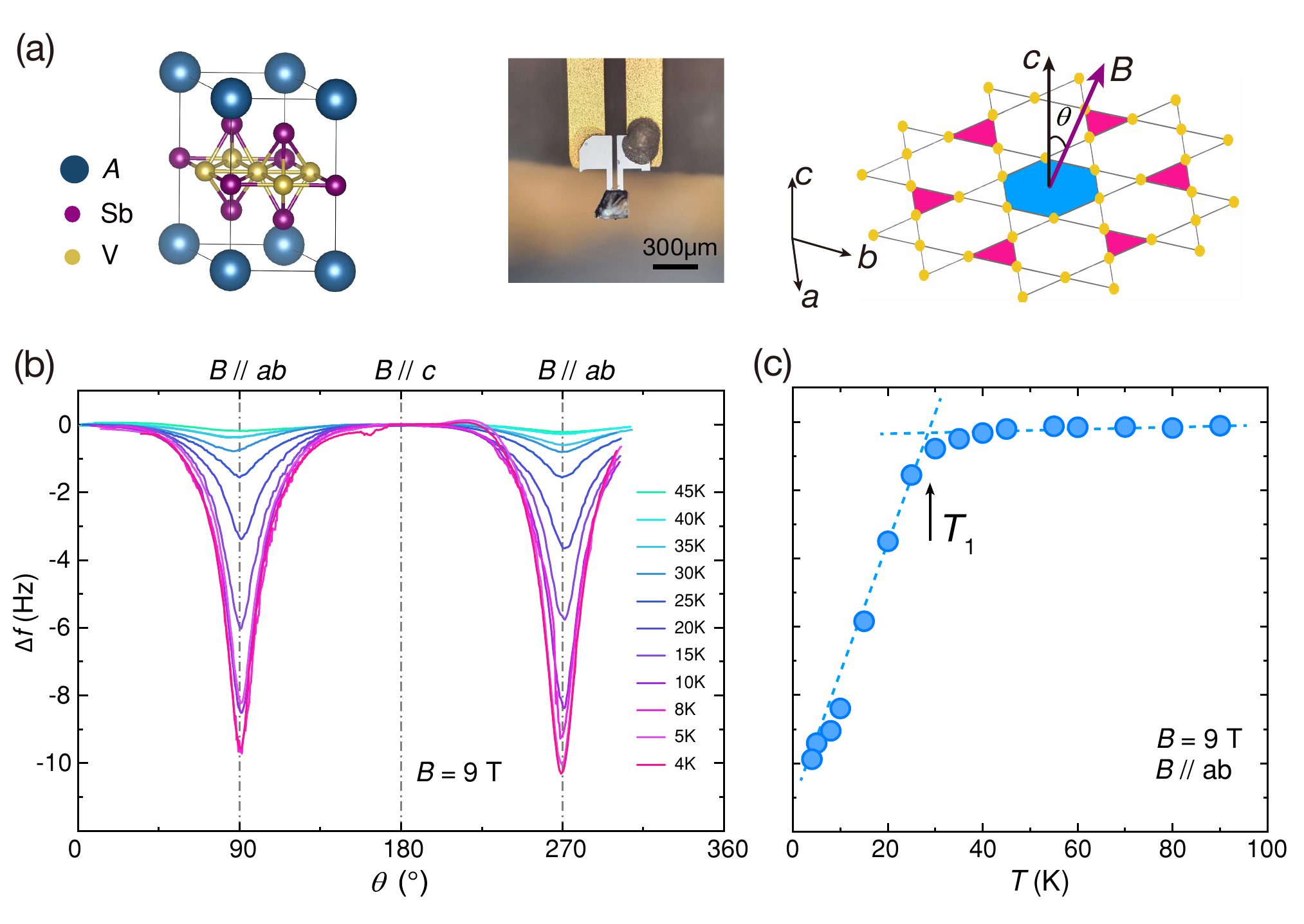}
\caption{ \textbf{Thermodynamic phase transition across 30~K.}
\textbf{a} Crystal structure of \CVS. The vanadium atoms form a kagome lattice in the V-Sb1 plane which is stacked between Cs and Sb2 layers. An optical image of the 3.4 micron-thick, 300 micron-long silicon lever with a 200 $\times$ 100 $\times$ 1 micron$^3$ \CVS\ crystal mounted at the tip. The sketch on the right highlights the vanadium kagome layer of \CVS\ and defines the notation for angle $\theta$ used in the work.
\textbf{b} Angular dependent relatively resonant frequency $\Delta f$ measured at various temperatures with a vertical field of 9~T. Sample (\#S1-1) is rotated out-of-plane at a sweep rate of $\omega_r$ = 0.05$^{\circ}$/sec.
\textbf{c} Averaged $\Delta f(T)$ with $\theta$ = 90$^{\circ}$ and 270$^{\circ}$ , i.e. $\textbf{B} \parallel ab$. }
\label{fig1}
\end{figure*}

We studied ultra high-quality single crystals which show a very large residual resistance ratio of 300-500 (see Supplementary Fig.~3). $\Delta f(\theta)$ ($\equiv$ $f(\theta)-f(180^{\circ})$) was measured from $c$-axis to the $ab$-plane up to 9~T, and in a temperature range from 20~mK to 110~K, see Fig.~\ref{fig1}(b) and Supplementary Fig.~4. When the magnetic response is in the linear regime, the angular dependence of magnetotropic susceptibility follows $\cos(2\theta)$ angular dependence~\cite{Arkady2023}. This is consistent with our high-temperature frequency data in Figure~\ref{fig1}(b) and Supplementary Fig.~4.
The tiny change of $f$ indicates the magnetic anisotropy is negligible even at 9~T, which is in line with the metallic and paramagnetic nature of \CVS. 

Below 30~K, $\Delta f$ quickly deviates from cos(2$\theta$) behavior and its magnitude increases dramatically (Fig.~\ref{fig1}(b)).
The sharp dips at 90$^{\circ}$ and 270$^{\circ}$, i.e. $\textbf{B} \parallel ab $, are signatures for extremely anisotropic magnetic moment along the $c$-axis, which could be nicely simulated by our simple model considering 2D ferromagnetic-type magnetic structure (see Supplementary Eq.~(6) and Supplementary Fig.~5.) Theoretically, the amplitude of $m$ is proportional to $\Delta f$ between $\textbf{B} \parallel ab$-plane and $c$-axis.
In Fig.~\ref{fig1}(c), we plot the temperature dependence of $\Delta f(\theta = 90^{\circ}, T)$. 
Indeed, the magnitude of $\Delta f(90^{\circ})$ exhibits a sharp cusp at 30~K, which increases by a factor of approximately 60 as we decrease temperature from 30~K to 4~K.
Such a sharp onset of the magnetotropic anisotropy below 30~K suggests a magnetic phase boundary. In this context, our measurement is very much consistent with the magnetic anomalies detected by non-linear conductivity, anomalous  Nernst effect, $\mu$SR, and STM~\cite{MollP2022,ZhaoZX2021,ChenDongPRB,ChenXH2022N}. In fact, results from different experiments can be scaled to the same behavior which points to a thermodynamic phase boundary at 30~K, see Supplementary Fig.~6. As the measured magnetization is small along both $c$-axis and $ab$-plane (see Supplementary Figs.~7 and 8), our observations indicate the new magnetic phase possesses a small but highly anisotropic magnetic moment. We noticed that a nematic phase (rotational-symmetry-breaking phase in the $ab$-plane) was observed around $T_1$ in previous studies~\cite{ChenXH2022N}. However, whether such an in-plane order could induce an out-of-plane magnet phase is an interesting question for further studies. In the current work, we cannot discern electronic nematicity directly.

\begin{figure*}[t!!] 
\includegraphics[width=0.85\textwidth]{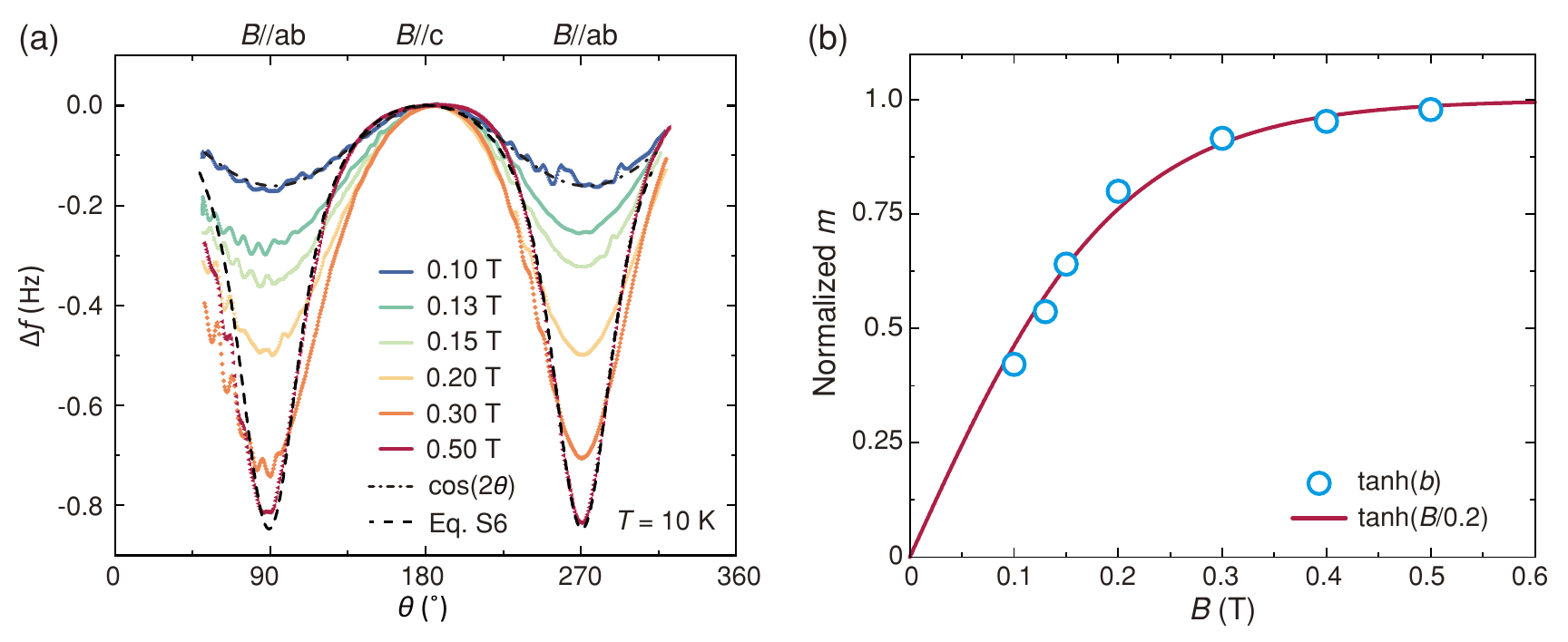}
\caption{\textbf{Evolution of resonant frequency at small magnetic field.}
\textbf{a} Angular dependence of the resonant frequency (sample \#S2-1) at 10~K for different small magnetic fields. 
\textbf{b} Comparison of a hyperbolic tangent magnetization function (tanh($B$/0.2), red line) and fitting parameter $b$ in Supplementary Eq.~(7) (blue circles). The three data points below 0.2~T are determined by fittings to cos(2$\theta$) behavior, which have relatively large error bars.}
\label{fig2}
\end{figure*}

To further investigate the novel magnetic phase, we conducted measurements on the magnetic field dependent $\Delta f(\theta)$ while maintaining the sample at 10~K. $\Delta f(\theta)$ is not only sensitive to temperature, but also shows a strong field dependence as demonstrated in Fig.~\ref{fig2}. 
At a relatively high magnetic field, above 0.3~T, $\Delta f(\theta)$ shares the same line-shape as those measured at 9~T below $\sim$30~K, see Fig.~\ref{fig2}(a). Upon reducing magnetic field, $\Delta f(\theta)$ reverts to a $\cos(2\theta)$ behavior below $\sim$0.15 T, providing evidence of the sample entering a linear regime. Figure~\ref{fig2}(b) compares the normalized magnetization curve in theory (by setting $B_c$ = 0.2~T) and the calculated \textit{m} based on Supplementary Eq.~(7) (with $b$ derived from the fittings). The obtained saturate field of 0.2~T is close to the value derived from non-linear conductivity (i.e. 0.15~T)~\cite{MollP2022}. This result indicates a small external magnetic field can saturate the magnetic moment along the $c$-axis of \CVS, or induce a new magnetic phase. Recently, C.Y. Guo et al., proposed that \CVS\ possesses “correlated order at the tipping point”~\cite{GuoNP2024}. In such case, a small strain or magnetic field may drive the sample away from its ground state and induce symmetry-broken phases. Such an understanding is also supported by recent theoretical work~\cite{Tazai2024}. The observed evolution of $\Delta f(\theta)$ curve in a small field (0.15~T) from our work is also consistent with the scenario of “correlated order at the tipping point”, and time reversal symmetry breaking phase exists only above certain magnetic field.

\begin{figure*}[t!!]
\includegraphics[width=0.95\textwidth]{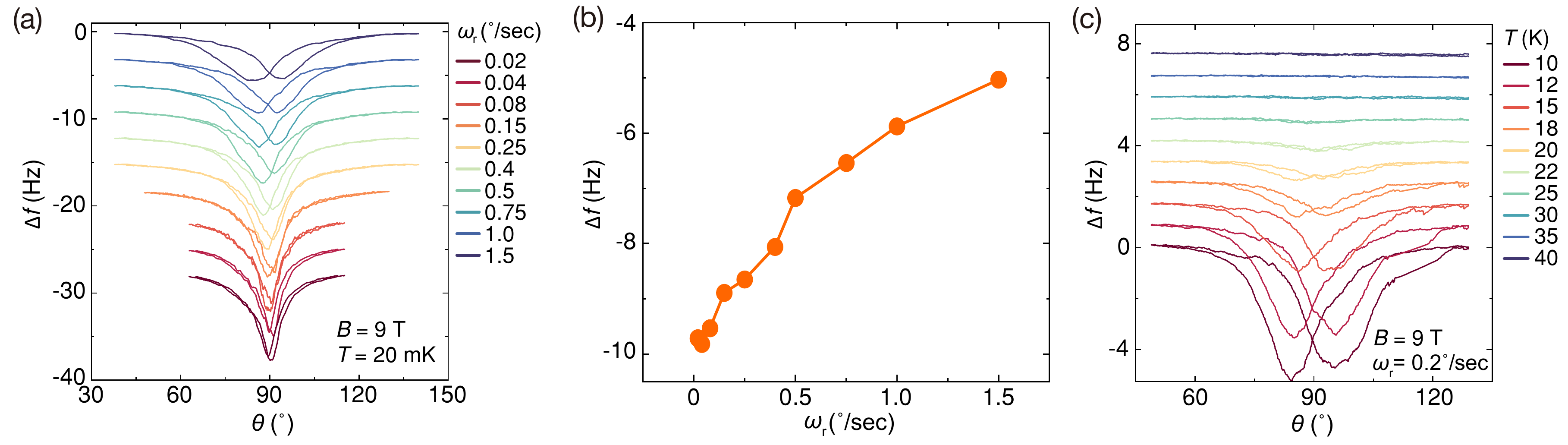}
\caption{\textbf{Slow dynamic of the low temperature magnetic phase.}
\textbf{a} Sweep rate ($\omega_r$) dependent $\Delta f(\theta)$ (sample \#S1-2) measured around 90$^{\circ}$ at 9~T and 20~mK. 
\textbf{b} $(\omega_r)$ dependent dip amplitude of $\Delta f(\theta)$ around 90$^{\circ}$. 
\textbf{c} Temperature dependent $\Delta f(\theta)$ (sample \#S1-1) measured in a magnetic field of 9~T, the sweep rate used here is 0.2$^{\circ}$/sec. For clarity, curves in \textbf{a} and \textbf{c} are offset vertically along down and up direction, respectively. }
\label{fig3}
\end{figure*}

Resonating frequency measurement also grants access to the relaxation dynamics of order parameters at very low frequencies~\cite{Arkady2023}. Our measurement at approximately 45~kHz---a very small scale compared to exchange interactions~\cite{Tazai2022} or other energy scales in \CVS---reveals the very slow dynamical characteristics of the magnetic phase below 30~K (Fig.~\ref{fig3}). When $\omega_r < 0.15^{\circ}$/sec, no clear hysteresis is discernible within our experimental resolution, as depicted in Fig.~\ref{fig3}(a). However, when measured at relatively high sweep rates ($\omega_r$ $>$ 0.25$^{\circ}$/sec), the resonant frequency displays hysteretic behavior (Fig.~\ref{fig3}(a)), illustrated by the splitting of the sharp dip at 90$^{\circ}$ between forward and backward sweep. Hysteresis typically emerges in magnetic materials due to time-reversal symmetry breaking together with the formation of domains. As the magnetic field rotates away from $\textbf{B} \parallel ab$, the component of the magnetic field along the $c$-axis ($B_c$) increases steadily. We assume that once $B_c$ surpass the coercive field, magnetic moments in different domains will be aligned along the external field.
Our data shows that, with increasing rotation speed, the split dips not only further separate with each other but also experience a decrease in their magnitude (Fig.~\ref{fig3}(b)). The reduced dip amplitude indicates the area between different magnetic domains tends to compensate. This hypothesis fits a picture of small domains form in \CVS\ at low temperature, which could be aligned by an external magnetic field, as proposed by Guo et al.,~\cite{MollP2022}.
Remarkably, the sweep rate of $0.15^{\circ}$/sec translates to more than 3 minutes between $60^{\circ}$ and $90^{\circ}$, highlighting an extremely sluggish relaxation dynamics of the domains. This is in contrast to conventional spin magnetism, where the time required to flip a spin is on the order of picoseconds~\cite{Sentef2022}. On the other hand, domain motion could take a long time.
Once sample temperature approaching 30~K, the hysteresis behavior is invisible as shown in Fig.~\ref{fig3}(c). Again, indicating 30~K is a new magnetic phase boundary.

\begin{figure}[t!!]
\includegraphics[width=0.8\linewidth]{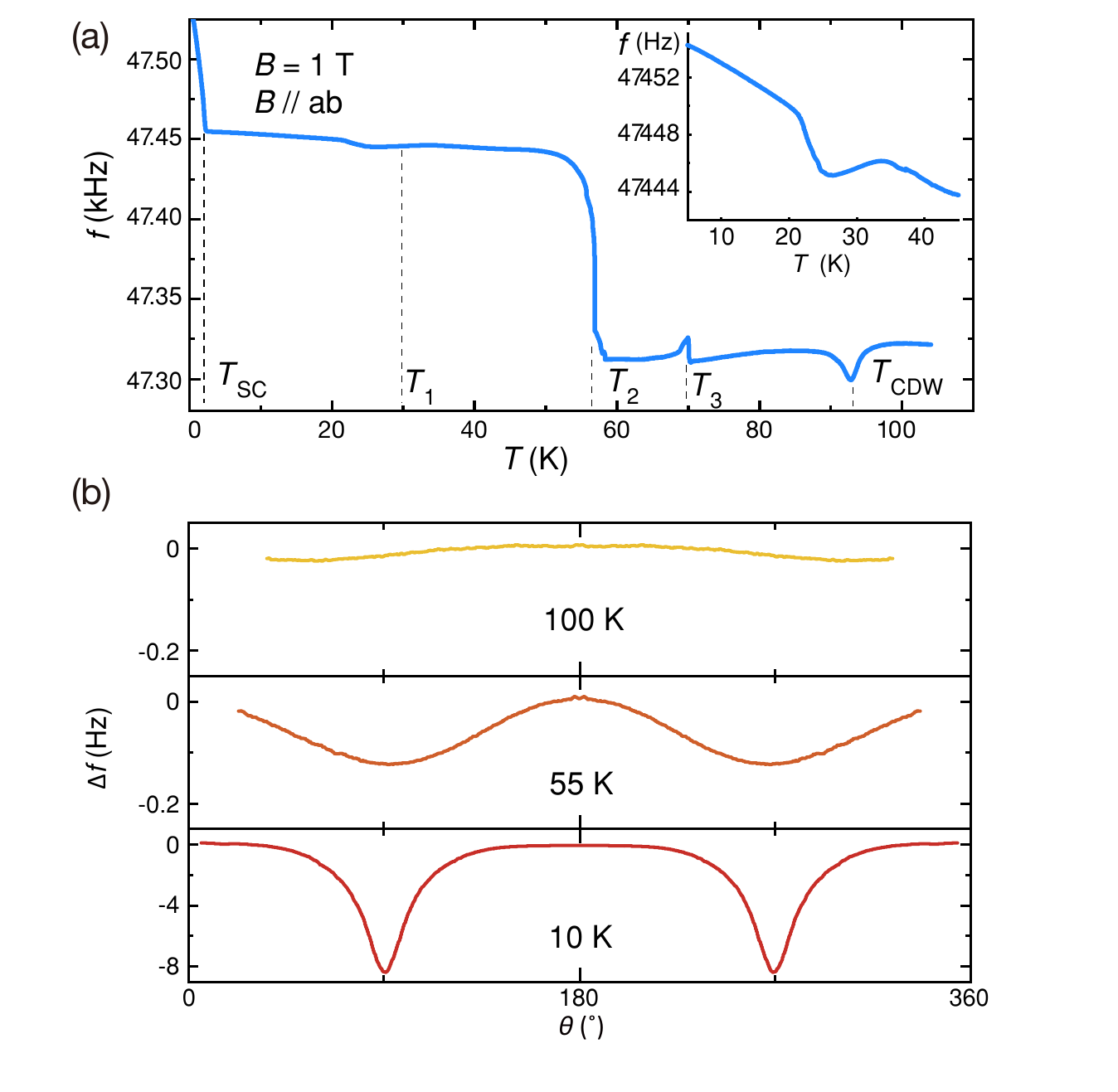}
\caption{\textbf{Evolution of the symmetry of the magnetotropic anisotropy.}
\textbf{a} Temperature dependent resonant frequency $f(T)$ (sample \#S1-1) from 110~K to 2~K. Several anomalies have been detected on $f(T)$ which are denoted in the figure. Inset shows the detail of $T_1$ transition. \textbf{b} Angular dependent $\Delta f(\theta)$ at selected temperatures. }
\label{fig4}
\end{figure}

To probe the evolution of the CDW state and to seek the origin of the low temperature magnetic phase, we contrast the behavior of the resonant frequency across the phase boundary at 30~K and its behavior across the well-established phase boundaries at high temperatures. Figure~\ref{fig4}(a) shows the resonant frequency at 1~T in the temperature range from 2~K to 100~K. In addition to a feature at $T_1$ = 30~K as discussed above, there is a series of anomalies being detected.
Abrupt drops in the resonant frequency at 94~K and 2~K are induced by the CDW and superconducting transitions, respectively.
In addition, we observed two additional anomalies: a sharp jump at 56~K, followed by a small dip at 70~K, ($T_2$ and $T_3$, respectively). The distinct jump at $T_2$ signifies a first-order phase transition, which is consistent with our high-resolution DC magnetization measurement, see Supplementary Figs.~3 and 8. X-ray diffraction reports a folding of the CDW superstructure along $c$-axis at $T_2$, i.e. from 2$\times$2$\times$4 to 2$\times$2$\times$2, just at $T_2$~\cite{GeckJ2022}. This suggests the 3D CDW order likely undergoes a change in its periodicity along the $c$-axis at $T_2$. $T_3$, however, is hardly distinguished through other transport or thermodynamic measurements.
Only zero-field $\mu$SR measurements suggest the onset of a time-reversal symmetry broken phase at 70~K~\cite{ZhaoZX2021}. 
This measurement also identified a large signal for time-reversal symmetry breaking below $T_1$. While $T_3$ was ascribed to a development of magnetic structure on two neighboring kagome layers with the out-of-plane magnetic coupling being anti-phase. Such a configuration results in a zero moment along $z$-axis and much weaker magnetotropic anisotropy between $T_1$ and $T_3$. 
There is another possibility that the quasiparticles quickly become coherent below 30-35 K. This has been suggested from STM study as the corresponding quasiparticle interference patterns are absent above this temperature but emerge with rotation symmetry breaking below it~\cite{ilija2023NP}.  
Likely, the establishment of such coherence enhances the orbital magnetic moment and allows us to observe the magnetotropic anisotropy below $T_1$.
This interpretation is consistent with the chiral magnetic effect observed below 30~K but not above~\cite{MollP2022}. 
On the other hand, if the time-reversal-symmetry breaking phase does occur above $T_1$ but without clear magnetic anisotropy in the $ac$-plane, it implies this phase is conjectured to have a three-dimensional structure, where local orbital magnetic moment contributions from adjacent layers cancel. Below $T_1$, different time-reversal-symmetry breaking states have been proposed (a staggered loop current ordered phase~\cite{ZhaoZX2021}, striped states with a unidirectional 4$a_0$ periodicity~\cite{Ilija2021N}, or chiral nematic states~\cite{Denner2021,Grandi2023}.) 
Nevertheless, the cascade of correlated electronic states in \CVS\ seems sensitive to sample quality and measuring environment. The high quality and small crystals used in our study might be crucial to reveal all these subtle phases.

Figure~\ref{fig4}(b) encompasses typical $\Delta f(\theta)$ at selected temperatures in a magnetic field of 9~T. There is a systematic evolution of the shape of $\Delta f(\theta)$ from 2~K to 120~K. Crossing $T_1$, $\Delta f(\theta)$ decays from sharp dips to a $\cos(2\theta)$ behavior as shown in Fig.~\ref{fig1}, which is a direct evidence for non-linear to linear magnetization response to the external magnetic field. Moreover, the cos(2$\theta$) behavior holds up to $T_{CDW}$ (see Supplementary Fig.~4 for details), which indicates that \CVS\ manifests itself as a paramagnetic metal with $\chi_{ab} > \chi_{c}$. The origin of this paramagnetic response root in the itinerant electrons which are partially gapped in the CDW phase~\cite{Gruner}. Above $T_{CDW}$, $\Delta f(\theta)$ does not show any sinusoidal behavior within our experimental resolution (upper panel), therefore the sample is either isotropic or possesses an exceedingly weak susceptibility. In other words, \CVS\ is a simple and good metal above $T_{CDW}$. We noticed that recent torque measurement in the $ab$-plane found the nematic phase even above $T_{CDW}$, together with evidence for time-reversal-symmetry breaking ~\cite{Matsuda_np_CsVSbtorque}. Although our torque measurement in the $ac$-plane detected signature for time-reversal-symmetry breaking phase below $T_1$ (Supplementary Fig.~9), our sample configurations (for both torque and tuning fork measurements) are not able to capture the in-plane nematic phase above~\cite{Matsuda_np_CsVSbtorque} or below $T_{CDW}$~\cite{ChenXH2022N}. Therefore, our experiment could not exclude possible in-plane time-reversal-symmetry breaking phase at high temperature.

\section{Discussions}
In the end, we discuss the possible origin of the magnetic phase below 30~K. The magnitude of magnetotropic susceptibility provides an opportunity to estimate the magnetic moment associated with magnetic structure in \CVS\ below 30~K. However, due to the small value of $\Delta f$ and a comparable background shift below 1~T, we could not precisely derive the absolute frequency corresponding to zero field ~\cite{Arkady2021} for different measuring conditions. Instead, we could only provide an upper limit of $m_c$ $\approx$ 0.03 $\mu_B$/vanadium ion (see Supplementary Note 2 and Supplementary Fig. 8 for details). Such a small magnetic moment is not consistent with electron-spin origin of the magnetism, and point toward orbital magnetism of itinerant electrons. It could also have been below the sensitivity limit of neutron scattering experiments~\cite{TobererES2019}. We note that a recent neutron scattering study derived an upper limit of orbital magnetic moment of $\sim$0.02$\pm$0.01$\mu_B$ per vanadium triangle~\cite{Neutron2024}, which is consistent with our observations.
Furthermore, our thermodynamic study unveiled the formation of domains below $T_1$  which may further reduce the size of the magnetic moment. With this regards, applying proper magnetic field, strain~\cite{GuoNP2024}, or even polarized photon (which might induce electro-striction and peizomagnetic effect)~\cite{Vidya_optical_RbVSb} can further break time-reversal or lattice mirror symmetry, resulting in a single domain and detectable moment.
Therefore, the small magnetic moment in our measurements is consistent with the picture of loop current CDW phase from experiments ~\cite{Hasan2021, MollP2022, ZhaoZX2021, Guguchia2022, WuL2022, Vidya_optical_RbVSb} and theories~\cite{HuJP2021,Balents2021,Nandkishore2021,NeupertT2021,Zhou2022,HuJP2021PRB}. Such a CDW phase can be illustrated by studying the kagome lattice tight-binding model described by nearest neighbor hoppings, see Supplementary Figs. 10-12 and Ref.~\cite{Zhou2022} for details. The loop current order is described by a combination of real and imaginary CDWs, the latter of which breaks the time reversal symmetry. While real CDW order parameters have been shown to be the leading instability in DFT calculations~\cite{Tan2021}, self-consistent Hartree-Fock studies have pointed out~\cite{NeupertT2021, Dong2023} that a CDW phase with both real and imaginary components can be stabilized by repulsive interactions beyond the onsite Hubbard term. Employing a simple mean-field calculation based on the LC2 state~\cite{Dong2023} (see Supplementary Section II for more details) as well as realistic estimates of the CDW order parameters lead to an orbital magnetic moment on the order of $\sim0.01\mu_B$ per vanadium atom. This outcome aligns consistently with the experimentally observed small magnetic moment $<$0.03$\mu_B$/V-atom, without invoking domain physics. We note that the extracted orbital magnetic moment is consistent with several time-reversal-symmetry breaking CDW patterns discussed in the literature~\cite{Denner2021}, and a more detailed analysis of which CDW pattern is stabilized is beyond the scope of this work.

In summary, our magnetotropic susceptibility measurements on \CVS\ have illuminated
a sequence of phase transitions below 100~K, which captures all the phase transitions reported in the literature.
By closely examining the line-shape of the angular dependent resonant frequency shifts ($\Delta f(\theta)$), we established that the phase transition at around 30~K breaks time-reversal symmetry by developing an orbital magnetic moment along the $c$-axis. The orbital magnetic moment saturates to a small value at low-temperatures and above 0.2~T, consistent with an itinerant
picture of magnetism due to staggered loop current. Therefore, our results provide novel insights that
significantly contribute to the comprehension of \AVS, thus shedding light on and
potentially resolving several ongoing debates within this dynamic field. Moreover, our discovery uncovers a magnetic structure that goes beyond the conventional framework of spin or atomic magnetism.

\cleardoublepage

\section{Methods}
\subsection{Sample preparation}
The \CVS~crystals were grown by self-flux method. Cs, V, and Sb with the atonic ratio of 7:3:14 were loaded in an alumina crucible, and was subsequently sealed in a tantalum tube and a quartz tube in turn. The tube was then heated to 1000~$^{\circ}$C, held for 20 hours, and then cooled to 400~$^{\circ}$C with a rate of 3~$^{\circ}$C/h before the furnace was turned off. The crystals were obtained after dissolving the flux by demineralized water. Samples have been carefully characterized as shown in the supplementary figures.

\subsection{Tuning fork measurement}
Part of the tuning fork measurements were conducted in Quantum Design Physical Property Measurement System (PPMS) system, the rest measurements were performed in $^3$He or dilution refrigerator systems in National High Magnetic Field Lab. The sample was placed at the tip of an Akiyama probe~\cite{Akiyamaprobe}. The resonance frequency of the probe was tracked by Zurich Instruments mid-frequency lock-in (MFLI) amplifier with the PLL/PID option. Detailed working principles and probe characterizations are included in the Supplementary Information.


\bibliographystyle{apsrev4-2}
\bibliography{CsV3Sb5.bib}

\cleardoublepage
\textbf{Acknowledgments} 
The authors thank Arkady Shekhter for his substantial help in data analysis and writing the manuscript. Work at Zhejiang University was supported by the National Key R\&D Program of China (Grant Nos. 2023YFA1406100 and 2022YFA1402200), the Zhejiang Provincial Natural Science Foundation of China (Grant No. LR25A040003), the Fundamental Research Funds for the Central Universities (Grants Nos. 226-2024-00039 and 226-2024-00068), and the National Natural Science Foundation of China (Grants Nos. 12374151, 12034017, and 12350710785). Measurements performed at the National High Magnetic Field Laboratory are supported by the National Science Foundation Cooperative Agreement (No. DMR-2128556, No. DMR-1644779) and the State of Florida. Z.W. is supported by the US DOE, Basic Energy Sciences Grant No. DE-FG02-99ER45747 and SEED Award No. 27856 from Research Corporation for Science Advancement. C.F., W. S., and D.C’s work at MPI-CPfS is supported by Deutsche Forschungsgemeinschaft (DFG) under SFB1143 (Project No. 247310070), the European Research Council Advanced Grant (No. 742068) “TOPMAT,” and the DFG through the Würzburg-Dresden Cluster of Excellence on Complexity and Topology in Quantum Matter ct.qmat (EXC 2147, Project-ID No. 39085490).

\textbf{Author Contributions Statement} 
L.J, X.W., and H.Y conceived the project. H.G., L.Y., Z.S., J.W., K.Z., and A.B. performed experimental measurements. D.C., W.S., and C.F. provided \CVS. J.Z, Y.L., and Y.Z. provided samples for control experiments. X.W., and Z.W. performed  theoretical analysis. L.J. and X.W. wrote the manuscript with the input of all authors.


\cleardoublepage

\cleardoublepage

\cleardoublepage

\cleardoublepage

\end{document}